\documentclass[journal]{IEEEtran}
\ifCLASSINFOpdf
  \usepackage[pdftex]{graphicx}
  \DeclareGraphicsExtensions{.pdf,.jpeg,.png}
\else
  \usepackage[dvips]{graphicx}
\fi
\usepackage{subfigure}
\usepackage{epstopdf}
\usepackage[cmex10]{amsmath}
\usepackage{amssymb}
\usepackage{wasysym}
\usepackage{algorithm}
\usepackage{algorithmic}
\usepackage{array}
\usepackage{cite}
\usepackage{color}
\usepackage{url}
\usepackage{subfigure}
\usepackage{epsfig,latexsym}%amsmath
\usepackage{flushend}
\usepackage{hyperref}
\usepackage{diagbox} % ¼ÓÔغê°ü
\usepackage{booktabs}
\usepackage{etoolbox}
\makeatletter
\begin{document}
%
% paper title
% can use linebreaks \\ within to get better formatting as desired
\title{Two-Step Codeword Design for Millimeter Wave Massive MIMO Systems with Quantized Phase Shifters}

% author names and affiliations
% use a multiple column layout for up to three different
% affiliations

% make the title area
\author{Kangjian~Chen,~\IEEEmembership{Student~Member,~IEEE}, Chenhao~Qi,~\IEEEmembership{Senior~Member,~IEEE}, \\ and Geoffrey Ye Li,~\IEEEmembership{Fellow,~IEEE}
\thanks{This work is supported in part by National Natural Science Foundation of China under Grant 61871119 and Natural Science Foundation of Jiangsu Province under Grant BK20161428. (\textit{Corresponding author: Chenhao~Qi})}
\thanks{Kangjian~Chen and Chenhao~Qi are with the School of Information Science and Engineering, Southeast University, Nanjing 210096, China (Email: qch@seu.edu.cn).}
\thanks{Geoffrey Ye Li is with the School of Electrical and Computer Engineering, Georgia Institute of Technology, Atlanta, GA, USA (Email: liye@ece.gatech.edu).}
}

\markboth{Accepted by IEEE Transactions on Signal Processing}
{Shell \MakeLowercase{\textit{et al.}}: Bare Demo of IEEEtran.cls for Journals}

\maketitle

\begin{abstract}
In this paper, a two-step codeword design approach for millimeter wave (mmWave) massive MIMO systems is presented. Ideal codewords are first designed, which ignores  the hardware constraints in terms of phase shifter resolution and the number of RF chains. Based on the ideal codewords, practical codewords are then obtained taking the hardware constraints into consideration. For the ideal codeword design in the first step, additional phase is introduced to the beam gain to provide extra degree of freedom. We develop a phase-shifted ideal codeword design (PS-ICD) method, which is based on alternative minimization with each iteration having a closed-form solution and can be extended to design more general beamforming vectors with different beam patterns. Once the ideal codewords are obtained in the first step, the practical codeword design problem in the second step is to approach the ideal codewords by considering the hardware constraints of the hybrid precoding structure in terms of phase shifter resolution and the number of RF chains. We propose a fast search based alternative minimization (FS-AltMin) algorithm that alternatively designs the analog precoder and digital precoder. Simulation results verify the effectiveness of the proposed methods and show that the codewords designed based on the two-step approach outperform those designed by the existing approaches.

\end{abstract}

\section{Introduction}
\begin{figure*}[htbp]
  \centering
  \includegraphics[width=0.9\textwidth]{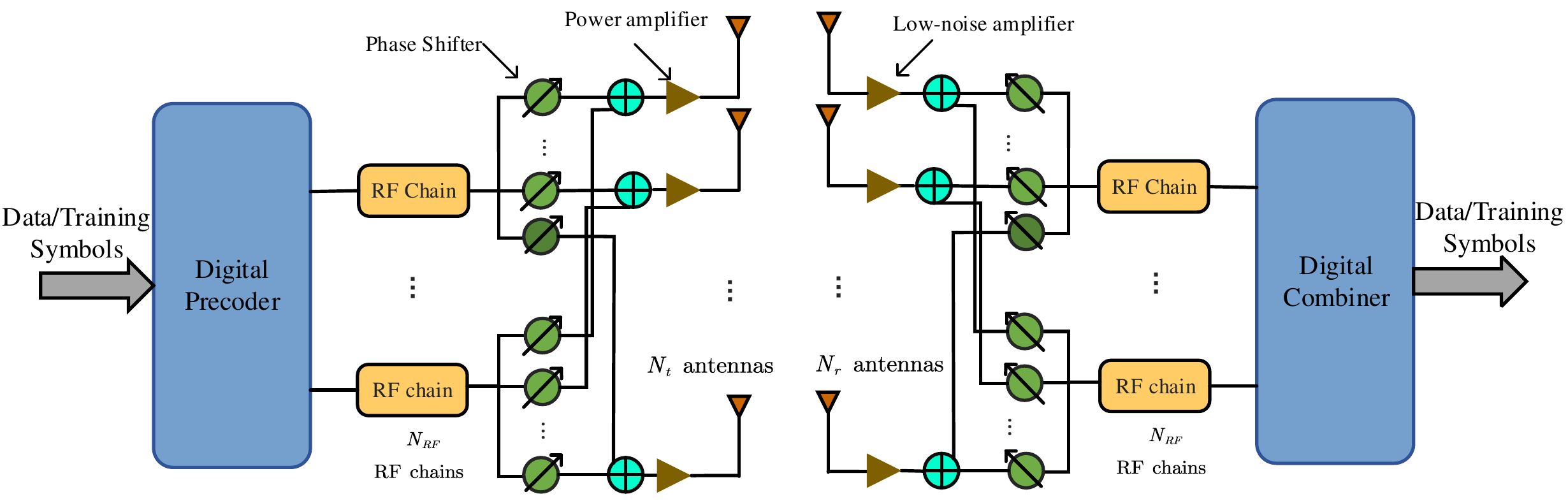}
  \caption{Illustration of a hybrid precoding and combing structure.}\label{fig:system model}
\end{figure*}

Millimeter wave (mmWave) communications have drawn extensive attention due to its rich spectrum resource to meet increasing demand in data traffic~\cite{RobertHeathOutdoorTSP2016,BLWandYeli,TSPCChen,TSPJchoi,LiYeEE,LiYemmWave}.  Its short wave length enables a large antenna array to be packed into a small area, which facilitates massive multi-input multi-output (MIMO) transmission to compensate the path loss induced by high frequency and support parallel transmission of data streams\cite{TSPMWY,ZhenyuXiao2018}. As shown in Fig.~\ref{fig:system model}, mmWave massive MIMO usually employs hybrid precoding where a small number of RF chains are connected to a large number of antennas~\cite{TSPXXue}. Hybrid precoding is typically a cascade of analog precoding and digital precoding. Analog precoding uses a phase shifter and achieves directional transmission, usually with constant envelop and limited resolution~\cite{Phaseshifter}. Digital precoding, which functions similarly as in the convention MIMO, is employed to mitigate the mutual interference among different data steams.

To acquire channel state information needed by hybrid precoding, beam training based on codebook is usually adopted, where the codebook is made up of a number of codewords that can simply implemented by channel steering vectors~\cite{CommonCodebookDesign,TSPMKok,TWCSXY}. To reduce the overhead of beam training, hierarchical codebook is introduced~\cite{Xiao2016Codebook}. Normally, a hierarchical codebook consists of a small number of low-resolution codewords covering wide angle at the upper layer of the codebook and a large number of high-resolution codewords offering high directional gain at the lower layer of the codebook\cite{Sparse2014}. Several hierarchical codebook design schemes have already been proposed~\cite{Sparse2014,PS-DFT2017,Xiao2016Hierarchical}. In~\cite{Sparse2014}, given the absolute beam gain as an objective, the least squares (LS) method is first applied to obtain an ideal codeword, which is generally named as LS ideal codeword design (LS-ICD). Based on the ideal codeword, the orthogonal matching pursuit (OMP) algorithm is used to obtain a practical codeword by considering the limited number of RF chains and the limited resolution of phase shifters. In~\cite{PS-DFT2017}, a phase-shifted discrete Fourier transform (PS-DFT) scheme is proposed, where the codewords are formed by the weighted summation of channel steering vectors. Although the hardware constraints in terms of RF chains and phase shifters are implicitly taken into account in the channel steering vectors, PS-DFT needs a large number of RF chains to achieve low-resolution codewords. In~\cite{Xiao2016Hierarchical}, a beam widening via single RF chain subarray (BMW-SS) is proposed, where the antenna array is divided into several sub-arrays and the codeword is designed via weighted summation of different beams formed by different sub-arrays.

% In \cite{JJZhang2017}, the beam design is formulated as an optimization problem, where the ripple in the main and side lobes of the beam is constrained. Although, hierarchical codebook with good property can be designed, the design method may be too complex.

In this paper, we present a two-step codeword design  approach for mmWave massive MIMO. In the first step, we design ideal codewords regardless of hardware constraints in terms of phase shifter resolution and the number of RF chains. We then design practical codewords based on ideal codewords by considering the hardware constraints in the second step. The main contribution of this paper is summarized as follows:
\begin{itemize}
	 \item  For the ideal codeword design, we introduce additional phase to the beam gain to provide extra degree of freedom and propose a phase-shifted ideal codeword design (PS-ICD) method. To determine the additional phase, an alternative minimization method is used, where each iteration of the method is based on a closed-form solution we derive. The proposed PS-ICD can also be extended to design more general beamforming vectors with different beam patterns.
	 \item Given the ideal codewords designed in the first step, the practical codeword design problem in the second step is to approach the ideal codewords by considering the hardware constraints of the hybrid precoding structure in terms of phase shifter resolution and the number of RF chains. We propose a fast search based alternative minimization (FS-AltMin) algorithm that alternatively designs the analog precoder and digital precoder.
\end{itemize}

The rest of this paper is organized as follows. The problem of codeword design is formulated in Section~\ref{Sec.Problem Formulation}. The ideal codeword design and practical codeword design are presented in Section~\ref{Sec.ICD} and Section~\ref{Sec.PCD}, respectively. Simulation results are provided in Section~\ref{Sec.Simulation}. Finally, Section~\ref{Sec.Conclusion} concludes this paper.

The notations used in this paper are defined as follows. Symbols for matrices (upper case) and vectors (lower case) are in boldface. $(\cdot)^T$, $(\cdot)^H $, $|\cdot |$, $\|\cdot \|_2$, $\mathbb{C}$, $\mathbb{R}$, $\mathbb{E}\{ \cdot \}$, ${\rm mod}(\cdot)$, $\mathcal{O}(\cdot)$, $\angle$, $\boldsymbol{I}$ and $\mathcal{C}\mathcal{N}$ denote the transpose, conjugate transpose (Hermitian), absolute value, $\ell_2$-norm, set of complex number, set of real number, operation of expectation, operation of modulo, order of complexity, angle, identity matrix and complex Gaussian distribution, respectively. $\left[ \boldsymbol{a} \right] _n$, $\left[ \boldsymbol{A} \right] _{n,:} $, $\left[ \boldsymbol{A} \right] _{:,m}$ and $\left[ \boldsymbol{A} \right] _{n,m}$ denote the $n$th entry of vector $\boldsymbol{a}$, the $n$th row of matrix $\boldsymbol{A}$, the $m$th column of matrix $\boldsymbol{A}$, and the entry on the $n$th row and $m$th column of matrix $\boldsymbol{A}$, respectively. ${\rm Re}\{\cdot\}$ and  ${\rm Im}\{\cdot\}$ denote the real part and imaginary part of a complex number, respectively.

\section{Problem Formulation}\label{Sec.Problem Formulation}
After briefly introducing mmWave  massive MIMO and beam training, in this section, we will formulate the problem of codeword design.
\subsection{MmWave Massive MIMO}
As shown in Fig.~\ref{fig:system model}, we consider a mmWave massive MIMO system with $N_{\rm t}$ and $N_{\rm r}$ antennas at the transmitter and receiver, respectively. Without loss of generality, we assume $N_{\rm t} \geq N_{\rm r}$. The antennas at both sides are placed in uniform linear arrays with half wavelength intervals. The mmWave massive MIMO system is equipped with the same number of $N_{\rm RF}$ RF chains at the transmitter and the receiver\cite{Sparse2014}. At the transmitter, each RF chain is fully connected to $N_{\rm t}$ antennas via quantized phase shifters, signal combiners, and power amplifiers. At the receiver, each RF chain is fully connected to $N_{\rm r}$ antennas via quantized phase shifters, signal combiners, and low-noise amplifiers. We will consider the phase shifters usually with limited resolution, e.g., six bits\cite{Phaseshifter}.

When transmitting $N_{\rm DT} (1\leq N_{\rm DT} \leq N_{\rm RF})$ data streams $\boldsymbol{x}\in \mathbb{C}^{N_{\rm DT}}$ in parallel, the received signal $\boldsymbol{y}\in \mathbb{C}^{N_{\rm DT}}$ can be expressed as
\begin{equation}\label{system model}
\boldsymbol{y}=\sqrt{P}\boldsymbol{W}_{\rm BB} ^{H}\boldsymbol{W}_{\rm RF}^{H}\boldsymbol{H}\boldsymbol{F}_{\rm RF} \boldsymbol{F}_{\rm BB} \boldsymbol{x} + \boldsymbol{W}_{\rm BB}^{H} \boldsymbol{W}_{\rm RF}^{H} \boldsymbol{\eta},
\end{equation}
where $\boldsymbol{F}_{\rm BB}\in \mathbb{C}^{N_{\rm RF}\times N_{\rm DT}}$, $\boldsymbol{F}_{\rm RF}\in \mathbb{C}^{N_{\rm t}\times N_{\rm RF}}$, $\boldsymbol{W}_{\rm BB}\in \mathbb{C}^{N_{\rm RF}\times N_{\rm DT}}$, $\boldsymbol{W}_{\rm RF}\in \mathbb{C}^{N_{\rm r}\times N_{\rm RF}}$ , $\boldsymbol{H}\in \mathbb{C}^{N_{\rm r}\times N_{\rm t}}$, and $\boldsymbol{\eta}\in \mathbb{C}^{N_{\rm r}}$ denote the digital precoder, the analog precoder, the digital combiner, the analog combiner, the mmWave MIMO channel matrix, and the additive white Gaussian noise vector with $\boldsymbol{\eta}\sim\mathcal{C}\mathcal{N}\left( \boldsymbol{0},\sigma _{\eta}^{2}\boldsymbol{I}_{N_{\rm r}} \right) $, respectively. Suppose the total power of the transmitter is $P$ and the transmit signal vector $\boldsymbol{x}$ is normalized such that $\mathbb{E}\left\{ \boldsymbol{x x}^H \right\} =\frac{1}{N_{\rm DT}}\boldsymbol{I}_{N_{\rm DT}}$. The hybrid precoder, including the digital precoder and the analog precoder has no power gain, i.e., $\|\boldsymbol{F}_{\rm RF}\boldsymbol{F}_{\rm BB}\|_2=1$. Similarly, the hybrid combiner, including the digital combiner and the analog combiner has no  power gain either, i.e., $\|\boldsymbol{W}_{\rm RF}\boldsymbol{W}_{\rm BB}\|_2=1$.

According to the widely used Saleh-Valenzuela channel model~\cite{heath2016overview}, the mmWave MIMO channel matrix $\boldsymbol{H}$ can be expressed as
\begin{equation}\label{channel model}
\boldsymbol{H}=\sqrt{\frac{N_{\rm t} N_{\rm r}}{L}}
\sum_{l=1}^L \mu_l\boldsymbol{a}(N_{\rm r},\Omega_l^{\rm r})\boldsymbol{a}(N_{\rm t},\Omega_l^{\rm t})^H,
\end{equation}
where $L$, $\mu_l$, $\Omega _{l}^{\rm r}$ and $\Omega _{l}^{\rm t}$ denote the number of multipath, the channel gain, the channel angle-of-arrival (AoA), and channel angle-of-departure (AoD) of the $l$th path, respectively. In fact, we have $\Omega _{l}^{\rm t}=\cos \left( \omega _{l}^{\rm t} \right)$ and $ \Omega_{l}^{\rm r}=\cos \left( \omega _{l}^{\rm r} \right)$, where $\omega _{l}^{\rm t}$ and $\omega _{l}^{\rm r}$ denote the physical AoA and AoD of the $l$th path, respectively. Since $\omega _{l}^{\rm t}\in \left[ 0,2\pi \right)$ and $\omega _{l}^{\rm r}\in \left[ 0,2\pi \right)$, we have $\Omega _{l}^{\rm t}\in [-1,1]$ and $\Omega _l^{\rm r} \in [-1,1]$. The channel steering vector $\boldsymbol{a}$ is defined as
\begin{equation}\label{steering vector}
\boldsymbol{a}\left( N,\Omega \right) =\frac{1}{\sqrt{N}}\left[ 1,e^{j\pi \Omega},\ldots ,e^{j\left( N-1 \right) \pi \Omega} \right] ^T,
\end{equation}
where $N$ is the number of antennas and $\Omega$ is the AoA or AoD.

\subsection{Beam Training}
Before the data transmission, the beam training tests all pairs of transmitting beam and receiving beam to find the pair best fit for the mmWave MIMO channel~\cite{Beamtrainning}. During the beam training stage, we only need to send a training symbol $x$ by setting $N_{\rm DT}=1$, which implies that we only need to use one column of $\boldsymbol{F}_{\rm BB}$ with all the other columns being zero. Therefore we can replace $\boldsymbol{F}_{\rm BB}\boldsymbol{x}$ by $\boldsymbol{f}_{\rm BB} x$ in \eqref{system model}. Correspondingly, to receive the training symbol, we only need to use one column of $\boldsymbol{W}_{\rm BB}$ with all the other columns being zero. Similarly, we can replace $\boldsymbol{W}_{\rm BB}$ by $\boldsymbol{w}_{\rm BB}$ in \eqref{system model}. Then
\begin{equation}\label{system model2}
y=\sqrt{P}\boldsymbol{w}^{H} \boldsymbol{H} \boldsymbol{v}x + \boldsymbol{w}^H \boldsymbol{\eta},
\end{equation}
where $\boldsymbol{v} \triangleq \boldsymbol{F}_{\rm RF} \boldsymbol{f}_{\rm BB}$ denotes the transmitting beam with the constraint $\| \boldsymbol{v}\|_2=1$ and $\boldsymbol{w} \triangleq \boldsymbol{W}_{\rm RF} \boldsymbol{w}_{\rm BB}$ denotes the receiving beam with the constraint $\| \boldsymbol{w}\|_2=1$. In fact, either $\boldsymbol{v}$ or $\boldsymbol{w}$ is essentially a codeword. The beam training aims at finding
the pair of $\boldsymbol{v}$ and $\boldsymbol{w}$  best fit for the mmWave MIMO channel, which can be expressed as
\begin{equation}\label{BeamTrainingOptimizationProblem}
  \arg\max_{\substack{\boldsymbol{v}\in \boldsymbol{V}_{\rm t}\\ \boldsymbol{w}\in \boldsymbol{V}_{\rm r}  }} \big| \boldsymbol{w}^H \boldsymbol{H} \boldsymbol{v}\big|,
\end{equation}
where $\boldsymbol{V}_{\rm t}$ and $\boldsymbol{V}_{\rm r}$ denote the codebook generated at the transmitter and the receiver, respectively. However, in practice it is impossible to find $\boldsymbol{v}$ and $\boldsymbol{w}$ directly from (5) due to the existence of the channel noise $\boldsymbol{\eta}$. We can only measure the power of $y$ to find the best pair of $\boldsymbol{v}$ and $\boldsymbol{w}$, which can be expressed as
\begin{equation}\label{BeamTrainingOptimizationProblem2}
  \arg\max_{\substack{\boldsymbol{v}\in \boldsymbol{V}_{\rm t}\\ \boldsymbol{w}\in \boldsymbol{V}_{\rm r}  }}  |y|^2.
\end{equation}
The solution of \eqref{BeamTrainingOptimizationProblem} and \eqref{BeamTrainingOptimizationProblem2} may be the same or different. If they are same, we call that the beam training is successful; otherwise, we say that the beam training is failed. The ratio of the number of successful beam training over the total number of beam training is defined as the success rate. Success rate of beam training is an important metric to evaluate $\boldsymbol{V}_{\rm t}$ and $\boldsymbol{V}_{\rm r}$.
\begin{figure}[!t]
  \centering
  \includegraphics[width=86mm]{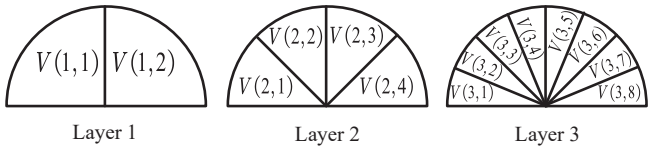}
  \caption{An example of hierarchical codebook with $M=2$ and $S=3$.}
  \label{fig:HierarchicalCodebook}
\end{figure}

In fact, a straightforward method to solve \eqref{BeamTrainingOptimizationProblem2} is to exhaustively search all pairs of $\boldsymbol{v}$ and $\boldsymbol{w}$ to find the best one. But such a method takes a long time for beam training. To improve the efficiency, beam training based on hierarchical codebooks is adopted~\cite{BeamsearchRobert,IrregularWaveform}. As shown in Fig.~\ref{fig:HierarchicalCodebook}, a hierarchical codebook typically consists of a small number of low-resolution codewords covering wide angle at the upper layer of the codebook and a large number of high-resolution codewords offering high directional gain at the lower layer of the codebook. The main feature of such a hierarchical codebook can be summarized as: $\boldsymbol{1)}$ The beams formed by the codewords in the same layer have the same width. $\boldsymbol{2)}$ The overall coverage of all codewords in the same layer is $[-1,1]$. $\boldsymbol{3)}$ Each beam formed by an upper layer codeword covers $M$ narrower beams formed by the lower layer codewords, where $M$ is called as hierarchical factor \cite{HierarchicalFactor}. Fig.~\ref{fig:HierarchicalCodebook} illustrates an example of hierarchical codebook $\boldsymbol{V}$ with $M=2$ and $S=3$. In general, $S$ is the layer of the codebook and is determined by $S=\lceil\log_M {N_{\rm t}}\rceil$. Compared to the exhaustive tests, the beam training for one data stream based on the hierarchical codebook can reduce the number of tests from $N_{\rm t}N_{\rm r}$ to $M\lfloor\log_M {N_{\rm t}}\rfloor+(M^2-M)\lfloor\log_M {N_{\rm r}}\rfloor$~\cite{Sparse2014}. For example, for an mmWave MIMO system with $N_{\rm t}=16$, $N_{\rm r}=8$ and $M=2$, the training overhead is reduced by $89\%$ . On the other hand, due to the low beam gain of the codewords at the upper layers of the hierarchical codebook, beam training may have lower success rate. Therefore, how to design a hierarchical codebook, especially the upper layer codewords is the focus of this paper.

Once the beam training is finished, $\boldsymbol{F}_{\rm RF}$ and $\boldsymbol{W}_{\rm RF}$ can be designed. Then we send pilot sequences to estimate the equivalent channel matrix which is a product of $\boldsymbol{F}_{\rm RF}$, $\boldsymbol{H}$ and $\boldsymbol{W}_{\rm RF}$. Once the equivalent channel matrix is estimated, we can design $\boldsymbol{F}_{\rm BB}$ and $\boldsymbol{W}_{\rm BB}$. Note that we do not directly estimate  $\boldsymbol{H}$. We only find the transmit beam and the receive beam best fit for the real channel matrix and then estimate the equivalent channel matrix~\cite{HierarchicalFactor}.

\subsection{Problem Formulation}
Note that the codebook is made up of a number of codewords. Therefore, in this paper, we focus on the codeword design. All the codewords in the codebook can be designed based on this work. In this paper, two important issues on codeword design are to be considered. One issue is the design of ideal codewords without consideration of hardware constraints in terms of phase shifter resolution and the number of RF chains. The other issue is the design of practical codewords based on the ideal codeword regarding the hardware constraint.

As in~\cite{CommonCodebookDesign} and \cite{Sparse2014}, the design of an ideal codeword $\boldsymbol{v}\in\mathbb{C}^{N_{\rm t}}$ with $\| \boldsymbol{v} \|_2=1$ commonly considers the following two objectives:
\begin{enumerate}
\item If the steering vector $\boldsymbol{a}(N_{\rm t},\Omega)$ is covered by $\boldsymbol{v}$, the absolute beam gain along the direction of $\Omega$ is a constant.
 \item  If $\boldsymbol{a}(N_{\rm t},\Omega)$ is not covered by $\boldsymbol{v}$, the beam gain is zero.
 \end{enumerate}

 The beam gain of $\boldsymbol{v}$ along $\Omega$ is denoted as a function of $\boldsymbol{v}$ and $\Omega$ as
\begin{equation}\label{beam gain}
G(\boldsymbol{v},\Omega)=\sqrt{N_{\rm t}}\boldsymbol{a}(N_{\rm t},\Omega)^H\boldsymbol{v}=\sum_{n=1}^{N_{\rm t}}[\boldsymbol{v}]_ne^{-j\pi(n-1)\Omega}
\end{equation}
for $\Omega\in[-1,1]$. Suppose the coverage of $\boldsymbol{v}$ is $\mathcal{I}_{v} \triangleq [\Omega_0,\Omega_0+B]$. Then from the aforementioned two objectives, we have
\begin{equation}\label{codeword obj}
    |G(\boldsymbol{v},\Omega)|=\left\{ \begin{array}{cl}
	C_v, &\Omega\in\mathcal{I}_{v}\\
	0,&\Omega\notin\mathcal{I}_v\\
\end{array} \right.
\end{equation}
where the absolute beam gain $C_v$ is determined by $\boldsymbol{v}$ and is independent of $\Omega$. This means that $C_v$ is the same for all possible $\Omega$ within the beam coverage $\mathcal{I}_v=[\Omega_0,\Omega_0+B]$.

Since it is impossible to find a codeword strictly satisfying \eqref{codeword obj}, we can only design ideal codewords that approach \eqref{codeword obj}~\cite{JJZhang2017}. We will introduce additional phase to $G(\boldsymbol{v},\Omega)$ to provide additional degree of freedom for ideal codeword design.

Once the ideal codewords are obtained, we can design practical codewords considering the hardware constraints in terms of phase shifter resolution and the number of RF chains~\cite{AltMin,SpatiallySparse,MulticastBeamDesign}. Given an ideal codeword $\boldsymbol{v}\in\mathbb{C}^{N_{\rm t}}$, the design of a practical codeword $\boldsymbol{v}_{\rm p} \triangleq \boldsymbol{\mathbf{F}}_{\rm RF}\boldsymbol{f}_{\rm BB} \in\mathbb{C}^{N_{\rm t}}$ is essentially to find $\boldsymbol{F}_{\rm RF}$ and $\boldsymbol{f}_{\rm BB}$, such that
\begin{subequations}\label{hybrid precoding}\normalsize
    \begin{align}
\underset{\boldsymbol{F}_{\rm RF},\boldsymbol{f}_{\rm BB}}{\min} &\|\boldsymbol{v}-\boldsymbol{F}_{\rm RF}\boldsymbol{f}_{\rm BB}\|_2 \label{Objective}\\
~~~\mathrm{s.t.} ~~~&\|\boldsymbol{F}_{\rm RF}\boldsymbol{f}_{\rm BB}\|_2=1, \label{Power Constraint}\\
    &\left[ \boldsymbol{F}_{\rm RF} \right] _{n,i}=e^{j \delta }, ~\forall \delta \in \boldsymbol{\Phi} _{b}, \label{envelop constrain}\\
    &n=1,2,\ldots,N_{\rm t},~i=1,2,\ldots, N_{\rm RF}, \nonumber
    \end{align}
\end{subequations}
where the constraint of \eqref{Power Constraint} indicates that the hybrid precoder provide no power gain, and the constraint of \eqref{envelop constrain} indicates each entry of the analog precoder satisfies the hardware restrictions of the phase shifters. If there are $b$ bits for quantized phase shifters, then
\begin{equation}\label{quantization bits}
\boldsymbol{\Phi}_{b}=\left[ \pi \left( -1+ \frac{1}{2^{b}}\right) ,\pi \left( -1+\frac{3}{2^{b}} \right) ,\ldots \pi \left( 1-\frac{1}{2^{b}} \right) \right].
\end{equation}

In the following, we will propose a two-step codeword design  approach for mmWave massive MIMO, which first designs ideal codewords regardless of hardware constraints and then designs practical codewords based on the ideal ones taking the hardware constraints into consideration.

\section{Ideal Codeword Design}\label{Sec.ICD}
In this section, as the first step of the two-step codeword design approach, we propose a phase-shifted ideal codeword design (PS-ICD) method. Additional phase is introduced to the beam gain to provide extra degree of freedom for ideal codeword design. To determine the additional phase, an alternative minimization method is used, where each iteration of the method is based on a closed-form solution we derive. The proposed PS-ICD can also be extended to design more general beamforming vectors with different beam patterns.

According to \eqref{codeword obj}, the beam coverage of an ideal codeword $\boldsymbol{v}\in\mathbb{C}^{N_{\rm t}}$ is $\mathcal{I}_v=[\Omega_0,\Omega_0+B]$, where we may temporarily omit the power constraint of $\| \boldsymbol{v} \|_2=1$ and assume $C_v=1$. Then \eqref{codeword obj} can be simplified as
\begin{equation}\label{g(Omega)codeowrd}
    |G(\boldsymbol{v},\Omega)|=\left\{ \begin{array}{cl}
	1, &\Omega\in\mathcal{I}_{v},\\
	0,&\Omega\notin\mathcal{I}_v.\\
\end{array} \right.
\end{equation}

Denote
\begin{equation}\label{GBD objective2}
  G(\boldsymbol{v},\Omega)=g(\Omega)e^{jf(\Omega)},
\end{equation}
where $g(\Omega)=|G(\boldsymbol{v},\Omega)|$ and $f(\Omega)=\angle G(\boldsymbol{v},\Omega)$ are both functions of $\Omega$. From \eqref{g(Omega)codeowrd}, we have
\begin{equation}\label{g(Omega)codeowrd2}
    g(\Omega)=|G(\boldsymbol{v},\Omega)| =\left\{ \begin{array}{cl}
	1, &\Omega\in\mathcal{I}_{v},\\
	0,&\Omega\notin\mathcal{I}_v.\\
\end{array} \right.
\end{equation}
Note that the condition in \eqref{g(Omega)codeowrd} has nothing to do with the phase $f(\Omega)$. Therefore, in our PS-ICD method, we introduce additional phase to increase the degree of freedom.

%We emphasize that the objective in \eqref{GBD objective2} with $f(\Omega)=0$ is exactly the motivation of LS-ICD in \cite{Sparse2014}.%

We define
\begin{equation}\label{matrix A}
\boldsymbol{A} \triangleq \sqrt{N_{\rm t}}[\boldsymbol{a}(N_{\rm t},\Omega_1),\boldsymbol{a}(N_{\rm t},\Omega_2),\ldots,\boldsymbol{a}(N_{\rm t},\Omega_K)],
\end{equation}
as a matrix made up of $K(K \geq N_{\rm t})$ channel steering vectors, where
\begin{equation}\label{quantization of Omega}
\Omega _k=-1+(2k-1)/{K},~k=1,2,\ldots,K,
\end{equation}
is the quantized channel AoD with equal interval. Note that the rank of $\boldsymbol{A}$ is $N_{\rm t}$.

To clearly represent $G(\boldsymbol{v},\Omega_k)$, for $k=1,2,\ldots,K $, we define a vector $\boldsymbol{g}\in\mathbb{C}^K$, with
\begin{equation}\label{entry of g}
  [\boldsymbol{g}]_k=g(\Omega_k)e^{jf(\Omega_k)}.
\end{equation}
%According to \eqref{beam gain}, \eqref{GBD objective2}, and \eqref{entry of g},
%\begin{equation}\label{objectives compact}
%\boldsymbol{A}^H\boldsymbol{v}=\boldsymbol{g},
%\end{equation}
%which is an over-determined equation.

The objective of the ideal codeword design is
\begin{equation}\label{CodewordDesignObjective}
  \underset{\boldsymbol{v},\boldsymbol{\Omega}}{\min}~\|\boldsymbol{A}^H\boldsymbol{v}-\boldsymbol{g}\|_2^2
\end{equation}
where $\boldsymbol{\Omega}=[f(\Omega_1),f(\Omega_2),\ldots,f(\Omega_K)]^T$. Given $\boldsymbol{{g}}$,  $\boldsymbol{\hat{v}}$ can be obtained by LS as
\begin{equation}\label{codeword LS}
\boldsymbol{\hat{v}}=\left( \boldsymbol{AA}^H \right) ^{-1}\boldsymbol{A{g}}.
\end{equation}
Direct calculation from \eqref{matrix A} yields $\boldsymbol{AA}^H=K\boldsymbol{I}_{N_{\rm t}}$. Then $\boldsymbol{\hat{v}}$ in \eqref{codeword LS} can be simplified as
\begin{equation}\label{hat_v with I}
  \boldsymbol{\hat{v}} = \frac{1}{K}\boldsymbol{Ag}.
\end{equation}
The objective in \eqref{CodewordDesignObjective} can be converted to
\begin{equation}\label{CodewordDesignObjective2}
  \underset{\boldsymbol{{\Omega}}}{\min}~\Big\|\frac{1}{K}\boldsymbol{A}^H\boldsymbol{Ag}-\boldsymbol{{g}}\Big\|_2^2.
\end{equation}
Note that
\begin{align}\label{ObjectiveAnalysis1}
  \Big\|\frac{1}{K}\boldsymbol{A}^H\boldsymbol{Ag}-\boldsymbol{{g}}\Big\|_2^2&=\Big(\frac{1}{K}\boldsymbol{g}^H\boldsymbol{A}^H\boldsymbol{A}-\boldsymbol{{g}}^H\Big)\Big(\frac{1}{K}\boldsymbol{A}^H\boldsymbol
  {Ag}-\boldsymbol{{g}}\Big)\nonumber\\
  &=\|\boldsymbol{{g}}\|_2^2-\frac{1}{K}\boldsymbol{g}^H\boldsymbol{A}^H\boldsymbol{A}\boldsymbol{g}.
\end{align}
Since $\|\boldsymbol{{g}}\|_2^2$ is predefined, we can further convert \eqref{CodewordDesignObjective2} to
\begin{equation}\label{CodewordDesignObjective3}
  \underset{\boldsymbol{\Omega}}{\max}~\boldsymbol{g}^H\boldsymbol{A}^H\boldsymbol{A}\boldsymbol{g}.
\end{equation}
From \eqref{CodewordDesignObjective3}, only the phases of $\boldsymbol{g}$ are involved. Without the constraint of \eqref{g(Omega)codeowrd2}, the solution of \eqref{CodewordDesignObjective3} is the eigenvector corresponding to the largest eigenvalue of $\boldsymbol{A}$. However, it is difficult to solve \eqref{CodewordDesignObjective3} with the constraint of~\eqref{g(Omega)codeowrd2}.

Since the phases of $\boldsymbol{g}$ are mutually independent, we can use the alternative minimization method to iteratively optimize each phase variable~\cite{AltMin,MulticastBeamDesign}. To be specific, we alternatively optimize $[\boldsymbol{\Omega}]_1,[\boldsymbol{\Omega}]_2,\ldots,[\boldsymbol{\Omega}]_K$ until a stop condition is satisfied.

When optimizing $[\boldsymbol{\Omega}]_k$ by fixing the other entries of $\boldsymbol{\Omega}$, i.e., $\{[\boldsymbol{\Omega}]_i,i=1,2,\ldots,K,i\neq k\}$, we can rewrite \eqref{CodewordDesignObjective3} as
\begin{align}\label{CodewordDesignObjective5}
\underset{\alpha,~\beta}{\max}~&\boldsymbol{t}^T\boldsymbol{R}\boldsymbol{t}\nonumber\\
  {\rm s.t.}~~&\alpha^2+\beta^2=g(\Omega_k)^2
\end{align}
 where
 \begin{equation}\label{denote1}
  \boldsymbol{R} \triangleq \left[\begin{matrix}
                   {\rm Re}\{\boldsymbol{A}^H\boldsymbol{A}\} & -{\rm Im}\{\boldsymbol{A}^H\boldsymbol{A}\} \\
                   {\rm Im}\{\boldsymbol{A}^H\boldsymbol{A}\} & {\rm Re}\{\boldsymbol{A}^H\boldsymbol{A}\}
                 \end{matrix}\right],~\boldsymbol{t} \triangleq \left[\begin{matrix}
                                                                 {\rm Re}\{\boldsymbol{g}\} \\
                                                                 {\rm Im}\{\boldsymbol{g}\}
                                                               \end{matrix}\right],
 \end{equation}
$\alpha \triangleq [\boldsymbol{t}]_k$ and $\beta \triangleq [\boldsymbol{t}]_{k+K}$.
Note that the optimization of $[\boldsymbol{\Omega}]_k$ is now converted as the optimization of real part and imaginary part of $[\boldsymbol{g}]_k$, i.e., $\alpha$ and $\beta$, respectively. We have
\begin{align}\label{ObjectiveAnalysis2}
  \boldsymbol{t}^T\boldsymbol{R}\boldsymbol{t} & = (\alpha[\boldsymbol{R}]_{k,:}+\beta[\boldsymbol{R}]_{k+K,:}+\boldsymbol{d}^T)\boldsymbol{t} \nonumber \\
   & = [\boldsymbol{R}]_{k,k}\alpha^2+[\boldsymbol{R}]_{k,k+K}\alpha\beta+p\alpha \nonumber\\
   &~~~+[\boldsymbol{R}]_{k+K,k+K}\beta^2+[\boldsymbol{R}]_{k+K,k}\alpha\beta\nonumber\\
   &~~~+q\beta+[\boldsymbol{d}]_k\alpha+[\boldsymbol{d}]_{k+K}\beta+r\nonumber\\
   & = [\boldsymbol{R}]_{k,k}\alpha^2+[\boldsymbol{R}]_{k+K,k+K}\beta^2+(p+[\boldsymbol{d}]_k)\alpha\nonumber\\
    &~~~+(q+[\boldsymbol{d}]_{k+K})\beta+([\boldsymbol{R}]_{k,k+K}+[\boldsymbol{R}]_{k+K,k})\alpha\beta+r \nonumber\\
    &\overset{(a)}{=}(p+[\boldsymbol{d}]_k)\alpha+(q+[\boldsymbol{d}]_{k+K})\beta\nonumber\\
    &~~~+([\boldsymbol{R}]_{k,k+K}+[\boldsymbol{R}]_{k+K,k})\alpha\beta+r+[\boldsymbol{R}]_{k,k}g(\Omega_k)^2\nonumber\\
    &\overset{(b)}{=}(p+[\boldsymbol{d}]_k)\alpha+(q+[\boldsymbol{d}]_{k+K})\beta+r+[\boldsymbol{R}]_{k,k}g(\Omega_k)^2
\end{align}
where
\begin{align}\label{ObjectiveAnalysis3}
   & \boldsymbol{d} \triangleq \left(\sum_{m\in\boldsymbol{\Psi}}~[\boldsymbol{t}]_m[\boldsymbol{R}]_{m,:}\right)^T,~r=\sum_{m\in\boldsymbol{\Psi}}~[\boldsymbol{d}]_m[\boldsymbol{t}]_m,\nonumber\\
   & p \triangleq \sum_{m\in\boldsymbol{\Psi}}~ [\boldsymbol{t}]_m[\boldsymbol{R}]_{k,m},~q \triangleq \sum_{m\in\boldsymbol{\Psi}}~ [\boldsymbol{t}]_m[\boldsymbol{R}]_{k+K,m}, \nonumber\\
   & \boldsymbol{\Psi}=\{1,2,\cdots,2K\}\backslash\{k,k+K\}
\end{align}
are independent of $\alpha$ and $\beta$, and are fixed once $[\boldsymbol{t}]_m~(m\in\boldsymbol{\Psi})$ is given. In \eqref{ObjectiveAnalysis2}, $(a)$ holds because $[\boldsymbol{R}]_{k,k}=[\boldsymbol{R}]_{k+K,k+K}=[{\rm Re}\{\boldsymbol{A}^H\boldsymbol{A}\}]_{k,k}$ and $\alpha^2+\beta^2=g(\Omega_k)^2$, while $(b)$ holds because $[\boldsymbol{R}]_{k,k+K}=[\boldsymbol{R}]_{k+K,k}=[{\rm Im}\{\boldsymbol{A}^H\boldsymbol{A}\}]_{k,k}=0$.
Then, \eqref{CodewordDesignObjective5} can be converted to
\begin{align}\label{CodewordDesignObjective6}
  \underset{\alpha,~\beta}{\max}~&(p+[\boldsymbol{d}]_k)\alpha+(q+[\boldsymbol{d}]_{k+K})\beta\nonumber\\
  {\rm s.t.}~~&\alpha^2+\beta^2=g(\Omega_k)^2.
\end{align}
As a result, the closed-form solution for \eqref{CodewordDesignObjective6} can be expressed as
\begin{align}\label{optimalsolution1}
&\alpha=\frac{g(\Omega_k)(p+[\boldsymbol{d}]_k)}{\sqrt{(p+[\boldsymbol{d}]_k)^2+(q+[\boldsymbol{d}]_{k+K})^2}},\nonumber\\
&\beta=\frac{g(\Omega_k)(q+[\boldsymbol{d}]_{k+K})}{\sqrt{(p+[\boldsymbol{d}]_k)^2+(q+[\boldsymbol{d}]_{k+K})^2}}.
\end{align}
We can obtain $[\boldsymbol{\Omega}]_k$ as
\begin{equation}\label{optimalsolution2}
[\boldsymbol{\Omega}]_k= \angle (\alpha+j\beta).
\end{equation}

Based on the steps from \eqref{CodewordDesignObjective5} to \eqref{optimalsolution2}, we iteratively optimize $[\boldsymbol{\Omega}]_1,[\boldsymbol{\Omega}]_2,\ldots,[\boldsymbol{\Omega}]_K$ until a stop condition is satisfied.

The stop condition is that the number of iterations equals  a predefined maximum number of iterations $R_{\rm max}$\footnote{Note that the ideal codeword design problem can also be treated as the design of FIR filter without special specifications~\cite{FIRDesign}, which essentially solves a series of convex optimization problems. Since our PS-ICD scheme has close-form solutions, it is much faster than solving convex optimization problems.}.

We denote $\boldsymbol{\hat{\Omega}}$ as the optimized $\boldsymbol{\Omega}$ after finishing the $R_{\rm max}$ iterations. We can obtain $\boldsymbol{\hat{g}}$ via \eqref{entry of g} and then obtain $\boldsymbol{\hat{v}}$ via \eqref{hat_v with I}. Since the power constraint of $\| \boldsymbol{v} \|_2=1$ is temporarily omitted in \eqref{g(Omega)codeowrd}, the designed ideal codeword will be
\begin{equation}\label{designed codeword2}
  \boldsymbol{v}=\frac{\boldsymbol{\hat{v}}}{\|\boldsymbol{\hat{v}}\|_2}.
\end{equation}

As shown in \textbf{Algorithm}~\ref{PV-ICD}, we summarize the steps of PS-ICD algorithm.

In fact, this part of work can be extended to design more general beamforming vectors with various beam patterns, which may be applied for beam training~\cite{IrregularWaveform}. Given a beam pattern $g(\Omega)$, the beamforming vector can be designed via \eqref{designed codeword2} with $\boldsymbol{g}$ obtained via alternative minimization algorithm. In Section~\ref{Sec.Simulation}, we will provide two examples to design general beamforming vectors, e.g., triangular beamforming vector and step beamforming vector.

\begin{algorithm}[!t]
	\caption{Phase-shifted Ideal Codeword Design (PS-ICD)}
	\label{PV-ICD}
	\begin{algorithmic}[1]
  \STATE \textbf{Input:} $g(\Omega)$,~$K$, $N_{\rm t}$, $R_{\rm max}$.
     \STATE  Randomly generate $\boldsymbol{\Omega}$.
    \STATE Set $i=1$.
            \WHILE{$i \leq R_{\rm max}$}
            \STATE Obtain  $k=\mod(i-1,K)+1$.
            \STATE Update $\boldsymbol{\Omega}$ via \eqref{optimalsolution2}.
            \STATE $i\leftarrow i+1$.
            \ENDWHILE
        \STATE Obtain $\boldsymbol{v}$ via \eqref{designed codeword2}.
       \STATE \textbf{Output:} $\boldsymbol{v}$.
	\end{algorithmic}
\end{algorithm}

\section{Practical Codeword Design}\label{Sec.PCD}
In this section, we will design practical codewords based on the ideal ones designed in Section~\ref{Sec.ICD}, by considering the hardware constraints in terms of phase shifter resolution and the number of RF chains. We will propose a fast search based alternative minimization (FS-AltMin) algorithm for the practical codeword design.
% We will divide the practical codeword design problem in \eqref{hybrid precoding} into several subproblems. For each subproblem, we will consider three different cases where the number of RF chains are one, two and more than two, respectively. With these findings, two algorithms including partial random search (PRS)-based and fast search (FS)-based practical codeword design algorithms will be proposed.
\cite{AltMin,SpatiallySparse}
In practice, we usually design $\boldsymbol{F}_{\rm RF}$ first, based on which we then design $\boldsymbol{f}_{\rm BB}$~\cite{Sparse2014}\cite{AltMin}. Therefore, we may temporarily ignore \eqref{Power Constraint} to design $\boldsymbol{F}_{\rm RF}$ since we can always satisfy \eqref{Power Constraint} by carefully adjusting $\boldsymbol{f}_{\rm BB}$. Then we have
\begin{subequations}\label{hybrid precoding2}\normalsize
    \begin{align}
\underset{\boldsymbol{F}_{\rm RF},\boldsymbol{f}_{\rm BB}}{\min} &\|\boldsymbol{v}-\boldsymbol{F}_{\rm RF}\boldsymbol{f}_{\rm BB}\|_2 \label{Objective2}\\
~~~~\mathrm{s.t.} ~~~ &\left[ \boldsymbol{F}_{\rm RF} \right] _{n,i}=e^{j \delta }, ~\forall \delta \in \boldsymbol{\Phi} _{b}, \label{envelop constrain2}\\
    &n=1,2,\ldots,N_{\rm t},~i=1,2,\ldots, N_{\rm RF}. \nonumber
    \end{align}
\end{subequations}
Although the OMP algorithm can be used to solve~\eqref{hybrid precoding2} and obtain a well fit for $\boldsymbol{v}$, it requires a large number of RF chains~\cite{Sparse2014,SpatiallySparse}, which occupies large hardware resource and reduce the energy efficiency. In this paper, we focus on how to solve~\eqref{hybrid precoding2} with a small number of RF chains compared to the OMP algorithm.

In the case of $N_{\rm RF}=1$, we can simply set $\boldsymbol{\widetilde{f}}_{\rm BB}=1/\sqrt{N_{\rm t}}$ to normalize the power and  set $[\boldsymbol{\widetilde{F}}_{\rm RF}]_n=e^{j\boldsymbol{[\widetilde{\theta}}]_n}$ where
\begin{equation}\label{One RF chain}
 [\boldsymbol{\widetilde{\theta}}]_n = \underset{[\boldsymbol{\boldsymbol{{\theta}}}]_n\in\boldsymbol{\Phi}_b, n=1,2,\cdots,N_{\rm t}}{\arg\min} \bigg|[\boldsymbol{\theta}]_n-\angle{[\boldsymbol{v}]_n}\bigg|.
\end{equation}
Note that $\boldsymbol{\widetilde{F}}_{\rm RF}$ is essentially reduced to be a vector in dimension, since there is only one RF chain. The entries of $\boldsymbol{\widetilde{F}}_{\rm RF}$ are mutually independent, indicating that there are limited degrees of freedom that can be explored for pracitcal codeword design.

In the case of $N_{\rm RF}\ge2$, since both $\boldsymbol{f}_{\rm BB}$ and $\boldsymbol{F}_{\rm RF}$ need to be designed, the problem in \eqref{hybrid precoding2} can be solved by an alternative minimization method, which can minimize \eqref{Objective2} by designing $\boldsymbol{f}_{\rm BB}$ with a fixed $\boldsymbol{F}_{\rm RF}$ and then designing $\boldsymbol{F}_{\rm RF}$ with a fixed $\boldsymbol{f}_{\rm BB}$~\cite{AltMin,MulticastBeamDesign}.

\subsection{Initialization}\label{Subsec.A}
First of all, an initial value is required to start the alternative minimization method. We may initialize either $\boldsymbol{F}_{\rm RF}$ or $\boldsymbol{f}_{\rm BB}$. Note that there are hardware constraints for $\boldsymbol{F}_{\rm RF}$ while there is no constraint for $\boldsymbol{f}_{\rm BB}$. We can initialize each entry of $\boldsymbol{F}_{\rm RF}$ by randomly selecting an entry from $\boldsymbol{\Phi}_b$.% To ease the notation, we denote the $i$th column of $\boldsymbol{F}_{\rm RF}$ as $\boldsymbol{f}_{i}$. Then \eqref{hybrid precoding2} can be rewritten as
\subsection{Design of $\boldsymbol{f}_{\rm BB}$ with a fixed $\boldsymbol{F}_{\rm RF}$} \label{Subsec.B}
Given a fixed $\boldsymbol{F}_{\rm RF}$, the problem in \eqref{hybrid precoding2} can be converted to
\begin{equation}\label{SolveoffBB}
  \underset{\boldsymbol{f}_{\rm BB}}{\min} \|\boldsymbol{v}-\boldsymbol{F}_{\rm RF}\boldsymbol{f}_{\rm BB}\|_2.
\end{equation}
It is a typical least squares problem with the  solution expressed as
\begin{equation}\label{ResultoffBB}
\boldsymbol{\widehat{f}}_{\rm BB}=(\boldsymbol{F}_{\rm RF}^H\boldsymbol{F}_{\rm RF})^{-1}\boldsymbol{F}_{\rm RF}^H\boldsymbol{v}.
\end{equation}

\subsection{Design of $\boldsymbol{F}_{\rm RF}$ with a fixed $\boldsymbol{f}_{\rm BB}$} \label{Subsec.C}
%However, it is challenging as the hybrid optimization problem involves variables $[\boldsymbol{f}_{\rm BB}]_i,i=1,2,\ldots,N_{\rm RF}$ as well as discrete variables $[\boldsymbol{f}_{i}]_n,i=1,2,\ldots,N_{\rm RF};n=1,2,\ldots,N_{\rm t}$.
Given a fixed $\boldsymbol{f}_{\rm BB}$ obtained via \eqref{ResultoffBB}, we can focus on the discrete optimization in terms of the quantized phase shifters. Then \eqref{hybrid precoding2} is converted into
%e.g., $\boldsymbol{\widehat{f}}_{\rm BB}$
\begin{subequations}\label{hybrid precoding omit2} \normalsize
  \begin{align}
\underset{\boldsymbol{F}_{\rm RF}}{\min}~ &\|\boldsymbol{v}-\boldsymbol{F}_{\rm RF}\boldsymbol{f}_{\rm BB}\|_2 \label{Objective3}\\
~~~~\mathrm{s.t.} ~ &\left[ \boldsymbol{F}_{\rm RF} \right] _{n,i}=e^{j \delta }, ~\forall \delta \in \boldsymbol{\Phi} _{b}, \label{envelop constrain3}\\
    &n=1,2,\ldots,N_{\rm t},~i=1,2,\ldots, N_{\rm RF}. \nonumber
    \end{align}
\end{subequations}
% Since each entry of $\boldsymbol{f}_{\rm sum}$ is a summation of different phases, the combinations of different phases resulting in the absolute value around $N_{\rm RF}/2$ appears much more frequently than those resulting in the absolute value of $N_{\rm RF}$. To provide more candidate phases for \eqref{hybrid precoding omit2} and ensure that we can find a proper solution to \eqref{hybrid precoding omit2}, we set
It is observed that the entries of the vector $(\boldsymbol{v}-\boldsymbol{F}_{\rm RF}\boldsymbol{f}_{\rm BB})$ are mutually independent, because the entries of $\boldsymbol{F}_{\rm RF}$ are independent to each other. Therefore, the minimization of $\| \boldsymbol{v}-\boldsymbol{F}_{\rm RF}\boldsymbol{f}_{\rm BB} \|_2$  can be converted to the minimization of the absolute value of each entry of $(\boldsymbol{v}-\boldsymbol{F}_{\rm RF}\boldsymbol{f}_{\rm BB})$. Then \eqref{hybrid precoding omit2} can be converted to $N_{\rm t}$ subproblems, where the $n(n=1,2,\ldots,N_t)$th subproblem can be expressed as
\begin{equation}\label{hybrid precoding each elment}
\begin{split}
\underset{[\boldsymbol{F}_{\rm RF}]_{n,:}}{\min} &\big|[\boldsymbol{v}]_n - [\boldsymbol{F}_{\rm RF}]_{n,:}\boldsymbol{f}_{\rm BB}\big| \\
~~~~\mathrm{s.t.} ~~ &\left[ \boldsymbol{F}_{\rm RF} \right] _{n,i}=e^{j \delta }, ~\forall \delta \in \boldsymbol{\Phi} _{b}, \\
    &~i=1,2,\ldots, N_{\rm RF}.
\end{split}
\end{equation}
Define
\begin{equation}\label{alphaebelta}
  \alpha_n e^{j\beta_n} \triangleq \left[\boldsymbol{v}\right] _n,
\end{equation}
where $\alpha_n \in [0,v_{\rm max}]$ and $\beta_n \in [-\pi,\pi)$ are the amplitude and the phase of $[\boldsymbol{v}]_n$, respectively. Since the method to solve \eqref{hybrid precoding each elment} is exactly the same for different $n$, we can omit the subscript $n$ and define $e^{j\theta_i} \triangleq [ \boldsymbol{F}_{\rm RF}]_{n,i}$. Then \eqref{hybrid precoding each elment} can be rewritten as
\begin{subequations}\label{elment reconstruct} \normalsize
\begin{align}
\underset{\theta _1,\theta_2,\ldots,\theta_{N_{\rm RF}}}{\min}\ &\bigg|\alpha_n e^{j\beta_n}-\sum_{i=1}^{N_{\rm RF}}{\left[ \boldsymbol{f}_{\rm BB} \right] _ie^{j\theta _i}}\bigg| \label{final_obj}\\
\mathrm{s.t.}~~~~~~~&\theta _i\in \boldsymbol{\Phi}_{b},~i=1,2,\ldots,N_{\rm RF}. \label{final_constraint}
\end{align}
\end{subequations}

To solve \eqref{elment reconstruct}, we consider two different cases, $N_{\rm RF}=2$ and $N_{\rm RF}>2$.

%\subsubsection{$N_{\rm RF}=1$}
%In case of $N_{\rm RF}=1$, there is only one variable $\theta_1$ in \eqref{elment reconstruct}, which can be rewritten as
%\begin{equation}\label{One RF chain}
 %\underset{\theta_1 \in \boldsymbol{\Phi}_{b}}{\min}\big|\alpha_n e^{j\beta_n}-e^{j\theta_1}\big|.
%\end{equation}
%The solution to \eqref{One RF chain} is denoted as $\widetilde{\theta}_1$, which is essentially to find an entry from $\boldsymbol{\Phi}_{b}$ closest to $\beta_n$.

\subsubsection{$N_{\rm RF}=2$}
In this case, there are two variables $\theta_1$ and $\theta_2$ in \eqref{elment reconstruct}, which can be rewritten as
\begin{equation}\label{two RF chain1}
 \underset{\theta_1 \in \boldsymbol{\Phi}_{b},~\theta_2 \in \boldsymbol{\Phi}_{b} }{\min}\big|\alpha_n e^{j\beta_n}-[\boldsymbol{f}_{\rm BB}]_1e^{j\theta_1} -[\boldsymbol{f}_{\rm BB}]_2e^{j\theta_2} \big|.
\end{equation}
Denote
\begin{equation}\label{fff}
\zeta_1e^{j\psi_1}\triangleq[\boldsymbol{f}_{\rm BB}]_1,~\zeta_2e^{j\psi_2}\triangleq[\boldsymbol{f}_{\rm BB}]_2
\end{equation}
where $\zeta_1$, $\psi_1$, $\zeta_2$ and $\psi_2$ are the amplitude of $[\boldsymbol{f}_{\rm BB}]_1$,  the phase of $[\boldsymbol{f}_{\rm BB}]_1$, the amplitude of $[\boldsymbol{f}_{\rm BB}]_2$ and the phase of $[\boldsymbol{f}_{\rm BB}]_2$, respectively.

We first ignore the constraints of $\theta_1,~\theta_2 \in \boldsymbol{\Phi}_{b}$ and solve
\begin{equation}\label{two RF chain2}
 \underset{\theta_1,\theta_2}{\min}\big|\alpha_n e^{j\beta_n}-\zeta_1e^{j\psi_1}e^{j\theta_1} -\zeta_2e^{j\psi_2}e^{j\theta_2} \big|.
\end{equation}
Therefore, we have $\alpha_n e^{j\beta_n}=\zeta_1e^{j\psi_1}e^{j\theta_1} +\zeta_2e^{j\psi_2}e^{j\theta_2}$, which is equivalent to
\begin{equation}\label{two RF chains eq}
\left\{ \begin{array}{ll}
\zeta_1\cos(\theta_1+\psi_1-\beta_n)+\zeta_2\cos(\theta_2+\psi_2-\beta_n)&=\alpha_n,\\
\zeta_1\sin(\theta_1+\psi_1-\beta_n)+\zeta_2\sin(\theta_2+\psi_2-\beta_n)&=0.\\
\end{array} \right.
\end{equation}
The solutions of \eqref{two RF chains eq} can be expressed as
\begin{equation}\label{two RF chains eq solution1}
\left\{ \begin{array}{l}
	\bar{\theta}_1=\beta_n-\psi_1+\arccos(\frac{\alpha_n^2+(\zeta_1+\zeta_2)(\zeta_1-\zeta_2)}{2\zeta_1\alpha_n}),\\
    \bar{\theta}_2=\beta_n-\psi_2-\arccos(\frac{\alpha_n^2-(\zeta_1+\zeta_2)(\zeta_1-\zeta_2)}{2\zeta_2\alpha_n}),\\
\end{array} \right.
\end{equation}
or
\begin{equation}\label{two RF chains eq solution2}
\left\{ \begin{array}{l}
\bar{\theta}_1=\beta_n-\psi_1-\arccos(\frac{\alpha_n^2+(\zeta_1+\zeta_2)(\zeta_1-\zeta_2)}{2\zeta_1\alpha_n}),\\
    \bar{\theta}_2=\beta_n-\psi_2+\arccos(\frac{\alpha_n^2-(\zeta_1+\zeta_2)(\zeta_1-\zeta_2)}{2\zeta_2\alpha_n}).\\
\end{array} \right.
\end{equation}
If we consider the constrains of $\theta_1,~\theta_2 \in \boldsymbol{\Phi}_{b}$,  then we can design $\theta_1$ and $\theta_2$ as
\begin{equation}\label{two RF chains result}
\left\{ \begin{array}{l}
	\widetilde{\theta} _1=\arg \underset{\theta_1 \in \boldsymbol{\Phi} _{b}}{\min}\ |\theta_1-\bar{\theta}_1 |,\\
	\widetilde{\theta} _2=\arg \underset{\theta_2 \in \boldsymbol{\Phi} _{b}}{\min}\ |\theta_2 -\bar{\theta}_2 |,\\
\end{array} \right.
\end{equation}
which is essentially to find two entries from $\boldsymbol{\Phi}_{b}$ closest to $\bar{\theta}_1$ and $\bar{\theta}_2$, respectively.
\subsubsection{$N_{\rm RF}>2$}
In case of $N_{\rm RF}>2$, there are more than two variables in \eqref{elment reconstruct}. If we first ignore the constraints in \eqref{final_constraint}
to solve \eqref{final_obj}, just as the case of $N_{\rm RF}=2$, the problem will be underdetermined, where there are more unknown variables than the equations. Although the method in~\cite{AltMin} can be used to solve the problem without the constraint in \eqref{final_constraint},  it can only obtain a solution with continuous phase.
When the solution is directly quantized, it suffers from the quantization errors.
%In this context, one method is to use the random search (RS), which repeatedly selects $N_{\rm RF}$ entries from $\boldsymbol{\Phi}_{b}$ as the value of
%$\theta_1,\theta_2,\ldots,\theta_{N_{\rm RF }}$ and figures out the objective of $|\alpha_n e^{j\beta_n}-\sum_{i=1}^{N_{\rm RF}}{e^{j\theta _i}}|$,
%and finally outputs the combination of $\theta_1,\theta_2,\ldots,\theta_{N_{\rm RF }}$ achieving the minimal objective. However, the computational complexity is very high and the convergence speed is too slow for such method.
Motivated by the case of $N_{\rm RF}=2$, we incorporate the calculation of \eqref{two RF chains result} into the case of $N_{\rm RF}>2$ and rewrite \eqref{elment reconstruct} as
\begin{align}\label{more than two RF chains Equatation1}
\underset{\theta _1,\theta_2,\ldots,\theta_{N_{\rm RF}}}{\min}\ &\bigg|\gamma_n  e^{j\phi_n}-\zeta_1e^{j\psi_1}e^{j\theta_1} -\zeta_2e^{j\psi_2}e^{j\theta_2}\bigg|\nonumber\\
\mathrm{s.t.}~~~~~~&\theta _i\in \boldsymbol{\Phi}_{b},~i=1,2,\ldots,N_{\rm RF}
\end{align}
where
\begin{equation}
    \gamma_n e^{j\phi_n}\triangleq\alpha_ne^{j\beta_n}-\sum_{i=3}^{N_{\rm RF}}\left[ \boldsymbol{f}_{\rm BB} \right] _i{e^{j{\theta} _i}}
\end{equation}
with $\gamma_n$ and $\phi_n$ representing the amplitude and the phase, respectively.

Given $\theta_3,\ldots,\theta_{N_{\rm RF }}$, e.g., selecting any $N_{\rm RF}-2$ entries from $\boldsymbol{\Phi}_{b}$ as ${\hat{\theta}_3},\ldots,{\hat{\theta}_{N_{\rm RF}}}$, we can find $\theta_1,~\theta_2 \in \boldsymbol{\Phi}_{b}$ by
\begin{equation}\label{more than two RF chains Equatation2}
 \underset{\theta_1,\theta_2\in\boldsymbol{\Phi}_{b}}{\min}\big|\hat{\gamma}_n e^{j{\hat{\phi}_n}}-\zeta_1e^{j\psi_1}e^{j\theta_1} -\zeta_2e^{j\psi_2}e^{j\theta_2} \big|
\end{equation}
where $\hat{\gamma}_n e^{j{\hat{\phi}_n}}\triangleq\alpha_ne^{j\beta_n}-\sum_{i=3}^{N_{\rm RF}}\left[ \boldsymbol{f}_{\rm BB} \right] _i{e^{j{\hat{\theta}} _i}}$, using the method as in the case of $N_{\rm RF}=2$. Our motivation is that we always leave two degrees of freedom such as $\theta_1$ and $\theta_2$, to best match the given $\theta_3,\ldots,\theta_{N_{\rm RF}}$ by solving \eqref{more than two RF chains Equatation2}. Denote the solution of \eqref{more than two RF chains Equatation2} as $\hat{\theta}_1,\hat{\theta}_2$.
With the given ${\hat{\theta}_3},\ldots,{\hat{\theta}_{N_{\rm RF}}}$ and the computed $\hat{\theta}_1,\hat{\theta}_2$ in \eqref{more than two RF chains Equatation2}, we can { compute} the following objective as
\begin{equation}\label{more than two RF chains error}
\mathcal{E}(\hat{\theta}_1,\hat{\theta}_2,\ldots,\hat{\theta}_{N_{\rm RF}})=\bigg|\alpha_n e^{j\beta_n}-\sum_{i=1}^{N_{\rm RF}}{[\boldsymbol{f}_{\rm BB}]_{i}e^{j\hat{\theta} _i}}\bigg|.
\end{equation}

From the above discussion, $\theta_1$ and $\theta_2$ can be computed for given ${{\theta}_3},\ldots,{{\theta}_{N_{\rm RF}}}$. Now \eqref{more than two RF chains Equatation1} can be converted into the following optimization problem as
\begin{align}\label{more than two RF chains error2}
 \underset{{\theta}_3,\ldots,{\theta}_{N_{\rm RF}}}{\min} &{\mathcal{E}({\theta}_1,{\theta}_2,\ldots,{\theta}_{N_{\rm RF}})} \nonumber\\
 \mathrm{s.t.}~~~~&{\theta} _i\in \boldsymbol{\Phi}_{b},~i=3,\ldots,N_{\rm RF}
\end{align}
where ${\theta}_1$ and ${\theta}_2$ can be determined with the given ${{\theta}_3},\ldots,{{\theta}_{N_{\rm RF}}}$ via \eqref{more than two RF chains Equatation2}. To obtain the solution to \eqref{more than two RF chains error2}, one straightforward method is the exhaustive search, which tests all the combinations of ${\theta}_3,\ldots,{\theta}_{N_{\rm RF}}$ and select the one with the minimum objective. Since each $\theta_i,i=3,\ldots,N_{\rm RF}$ has $2^b$ possibilities, the exhaustive search needs totally $2^{(N_{\rm RF}-2)b}$ tests to obtain the solution to \eqref{more than two RF chains error2}. For example, if $N_{\rm RF}=6$ and $b=6$, we need totally $2^{24}$ tests, which has prohibitively high computation. In this context, sub-optimal search methods are usually adopted to find an appropriate solution to \eqref{more than two RF chains error2}.

Now we propose a fast search (FS) method for the design of $\boldsymbol{F}_{\rm RF}$ based on $\boldsymbol{\widehat{f}}_{\rm BB}$. The detailed steps are summarized in \textbf{Algorithm~1}. At step~8, we initialize ${\theta} _{3},\ldots ,{\theta} _{N_{\rm RF}}$ to be ${\theta} _{3}^{\rm 0},\ldots ,{\theta} _{N_{\rm RF}}^{\rm 0}$, {where the superscript ``0'' represents the number of iterations}. In fact, ${\theta} _{3}^{\rm 0},\ldots ,{\theta} _{N_{\rm RF}}^{\rm 0}$ can be set as the corresponding entries of the last computed $\boldsymbol{F}_{\rm RF}$ to guarantee the objective of \eqref{hybrid precoding2} is always decreasing.

At the $t(t\geq 1)$th iteration, we determine the value of $\theta_p(p=3,\ldots N_{\rm RF})$ as follows, where
\begin{equation}\label{the variable to be computed}
p=\mathrm{mod}\left( t-1,N_{\rm RF}-2\right) +3.
\end{equation}
Note that \eqref{the variable to be computed} essentially guarantees that the $N_{\rm RF}-2$ variables are computed cyclically.

We keep all the entries except $\theta_p^{(t)}$ to be the same as those in the $(t-1)$th iteration, which can be  expressed as
\begin{equation} \label{3toN_RF}
\theta _{i}^{(t)}=\theta _{i}^{(t-1)}, i=3,\ldots,N_{\rm RF}, i\neq p.
\end{equation}
Then given these $N_{\rm RF}-3$ entries, we test all entries from $\boldsymbol{\Phi} _{b}$ to determine $\theta_p^{(t)}$. For each test, given an entry from $\boldsymbol{\Phi} _{b}$, denoted as $\hat{\theta}_p \in \boldsymbol{\Phi} _{b}$, we obtain $\hat{\theta}_1$ and $\hat{\theta}_2$ via \eqref{more than two RF chains Equatation2} and then compute $\mathcal{E}(\hat{\theta}_1,\hat{\theta}_2,\theta_3^{(t)},\ldots,\theta_{p-1}^{(t)},\hat{\theta}_p, \theta_{p+1}^{(t)}, \ldots, \theta_{N_{\rm RF}}^{(t)})$. From all of these tests, we find a best $\hat{\theta}_p \in \boldsymbol{\Phi} _{ _{b}}$ satisfying
\begin{equation}\label{current variables}
\min_{\substack{\hat{\theta}_p \in \boldsymbol{\Phi} _{b}}} \mathcal{E}(\hat{\theta}_1,\hat{\theta}_2,\theta_3^{(t)},\ldots,\theta_{p-1}^{(t)},\hat{\theta}_p, \theta_{p+1}^{(t)}, \ldots, \theta_{N_{\rm RF}}^{(t)}).
\end{equation}
The solution to \eqref{current variables} is denoted as $\theta _{p}^{(t)}$.

We iteratively perform these steps until the stop condition expressed as
\begin{equation}\label{StopCondition}
  \theta_i^{(t)}= \theta_i^{(t+3-N_{\rm RF})},~i=3,4,\ldots, N_{\rm RF}
\end{equation}
is satisfied. In fact, \eqref{StopCondition} means the exactly same routine of the iterations is repeated again, which indicates the results thereafter will keep the same.
Once is \eqref{StopCondition} satisfied, we obtain
\begin{equation}\label{FinalSetValue}
  \widetilde{\theta}_i=\theta_i^{(t)},~i=1,2,\ldots, N_{\rm RF}.
\end{equation}

Then the $n(n=1,2,\ldots,N_{\rm t})$th row of the designed $\boldsymbol{F}_{\rm RF}$ is
$\boldsymbol{\widehat{F}}_{\rm RF}$ as
\begin{equation}\label{F_RF_row}
  [\boldsymbol{\widehat{F}}_{\rm RF}]_{n,:}=\left[ e^{j\widetilde{\theta}_{1}},e^{j\widetilde{\theta}_{2}},\ldots ,e^{j\widetilde{\theta}_{N_{\rm RF}}} \right].
\end{equation}

Finally, we output $\boldsymbol{\widehat{F}}_{\rm RF}$.

\begin{algorithm}[!t]
	\caption{FS Method for $\boldsymbol{F}_{\rm RF}$ Design}
	\label{alg1}
	\begin{algorithmic}[1]
        \STATE \textbf{Input:} $N_{\rm RF}$, $b$, $\boldsymbol{v}$, $\boldsymbol{\widehat{f}}_{\rm BB}$.
        \STATE Obtain $\boldsymbol{\Phi}_{b}$  via \eqref{quantization bits}.
        \FOR{$n=1,2,\ldots,N_{\rm t}$}
        \IF{$N_{\rm RF}=2$}
            \STATE Obtain $\widetilde{\theta}_1$ and $\widetilde{\theta}_2$ via \eqref{two RF chains result}.
        \ELSE
            \STATE Set $t=1$.		
            \STATE Initialize $\theta _{3}^{0},\theta _{4}^{0},\ldots ,\theta _{N_{\rm RF}}^{0}$.
            \WHILE{\eqref{StopCondition} is not satisfied}
            \STATE Obtain $\theta _{3}^{(t)},\theta _{4}^{(t)},\ldots ,\theta _{N_{\rm RF}}^{(t)}$ via \eqref{3toN_RF} and \eqref{current variables}.
            \STATE $t \leftarrow t+1$.
            \ENDWHILE
             \STATE Obtain $\widetilde{\theta} _{1},\widetilde{\theta}_{2},\ldots,\widetilde{\theta}_{N_{\rm RF}}$ via \eqref{FinalSetValue}.
        \ENDIF
        \STATE Obtain $[{\boldsymbol{\widehat{F}}}_{\rm RF}]_{n,:}$ via \eqref{F_RF_row}.
        \ENDFOR
       \STATE \textbf{Output:} $\boldsymbol{\widehat{F}}_{\rm RF}$.
	\end{algorithmic}
\end{algorithm}

\subsection{Design of practical codeword}
Now we propose the FS-AltMin algorithm for practical codeword design. Note that the purpose to introduce the name of practical codeword is to be consistent with the ideal codeword in concept. \textbf{Algorithm~2} summarizes the Subsection~\ref{Subsec.A}, Subsection~\ref{Subsec.B} and Subsection~\ref{Subsec.C}. We iteratively run the algorithms in Section IV-B and Section IV-C  to obtain $\boldsymbol{f}_{\rm BB}$ and $\boldsymbol{F}_{\rm RF}$ until the stop condition is satisfied. We may set the stop condition as that a predefined maximum number of iterations $T_{\rm max}$ is achieved or the entries in $\boldsymbol{f}_{\rm BB}$  do not change any more.

We denote the obtained $\boldsymbol{{F}}_{\rm RF}$ and $\boldsymbol{f}_{\rm BB}$  by the FS-AltMin algorithm as $\boldsymbol{\overline{F}}_{\rm RF}$ and $\boldsymbol{\overline{f}}_{\rm BB}$, respectively.  Note that in \eqref{hybrid precoding2}  we have temporarily ignored the power normalization constraint to simplify the practical codeword design. If the power normalization constraint in  \eqref{Power Constraint} is considered, we set
\begin{equation}\label{FRF fb opt}
\boldsymbol{\widetilde{f}}_{\rm BB}=\frac{\boldsymbol{\overline{f}}_{\rm BB}}{\|{\boldsymbol{\overline{F}}}_{\rm RF}\boldsymbol{\overline{f}}_{\rm BB}\|_2}.
\end{equation}
Finally, the designed practical codeword is
\begin{equation}\label{practical codeword}
  \boldsymbol{v}_{\rm p}=\boldsymbol{\overline{\boldsymbol{F}}}_{\rm RF}\boldsymbol{\widetilde{f}}_{\rm BB}.
\end{equation}

\subsection{Complexity Analysis}
It can be observed that the FS-AltMin algorithm needs at most $T_{\rm max}$ iterations. During each iteration, $N_{\rm t}$ subproblems expressed in \eqref{elment reconstruct} need to be solved. For each subproblem, we consider two cases. If $N_{\rm RF}=2$, we can get a solution directly via \eqref{two RF chains result} without any iterations. If $N_{\rm RF}>2$, we denote the number of iterations to achieve the convergence in \eqref{StopCondition} as $N_{\rm iter}$, where we need to compute \eqref{more than two RF chains error} for $2^b$ times in each iteration. Therefore, the computational complexity in solving \eqref{hybrid precoding2} is $\mathcal{O}(N_{\rm t}T_{\rm max})$ or $\mathcal{O}(N_{\rm t}T_{\rm max}N_{\rm iter}2^b)$ for $N_{\rm RF}=2$ or $N_{\rm RF}>2$, respectively.

\begin{algorithm}[!t]
	\caption{FS-AltMin algorithm for Practical Codeword Design}
	\label{alg2}
	\begin{algorithmic}[1]
        \STATE \textbf{Input:} $T_{\rm max}$, $\boldsymbol{v}$, $b$, $N_{\rm RF}$.
        \STATE Obtain $\boldsymbol{\Phi}_{b}$  via \eqref{quantization bits}.
        \STATE Initialize ${\boldsymbol{F}}_{\rm RF}$.
        \WHILE{\textit{Stopping Condition} is not satisfied}
        \STATE Obtain $\boldsymbol{\hat{f}}_{\rm BB}$ via \eqref{ResultoffBB}.
        \STATE Obtain ${\boldsymbol{\hat{F}}}_{\rm RF}$ via \textbf{Algorithm 1}.
         \ENDWHILE
        \STATE Obtain $\boldsymbol{v}_{\rm p}$ via \eqref{practical codeword}.
       \STATE \textbf{Output:} $\boldsymbol{v}_{\rm p}$.
	\end{algorithmic}
\end{algorithm}

\section{Simulation Results}\label{Sec.Simulation}
Now we evaluate the performance of the proposed two-step codeword design approach for mmWave massive MIMO systems with quantized phase shifters. The ideal codewords are designed with $K=128$ and $R_{\max}=2000$.

\begin{figure}[!t]
\centering
\includegraphics[width=88mm]{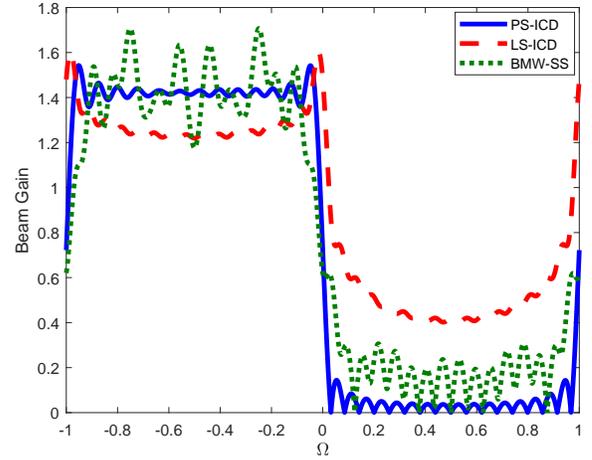}
\caption{Comparison of the beam patterns for different ideal codeword design methods.}
\label{fig:beam pattern}
\end{figure}

Fig.~\ref{fig:beam pattern} compares the beam patterns of ideal codewords designed by PS-ICD, BMW-SS~\cite{Xiao2016Hierarchical} and LS-ICD~\cite{Sparse2014} with $N_{\rm t}=32$ and $\mathcal{I}_v=[-1,0]$. From the figure, PS-ICD is better than BMW-SS and LS-ICD since the codeword designed by PS-ICD is flatter in main lobe and has larger attenuation in side lobe. Table~\ref{Tab.max deviation} compares the main lobe variation for different ideal codewords design methods by computing mean-squared errors (MSE) between the designed codewords and the objective in \eqref{codeword obj}. $C_v=\sqrt{2}$ is set to guarantee that the power of codewords is normalized, which can be obtained from Lemma 1 of ~\cite{CommonCodebookDesign}. From the table, for different $N_{\rm t}$, PS-ICD has the smallest variation among the three methods.

\begin{figure}[!t]
\centering
\subfigure[Comparison of triangular beam patterns for PS-ICD and LS-ICD.]{
\begin{minipage}[b]{0.5\textwidth}
\includegraphics[width=87mm]{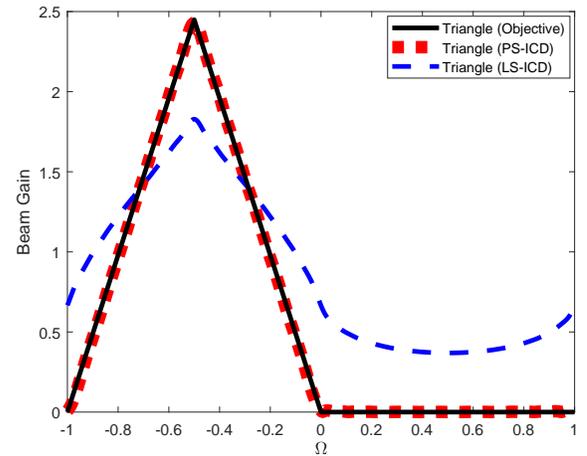}
\end{minipage}
}
\subfigure[Comparison of step beam patterns for PS-ICD and LS-ICD.]{
\begin{minipage}[b]{0.5\textwidth}
\includegraphics[width=87mm]{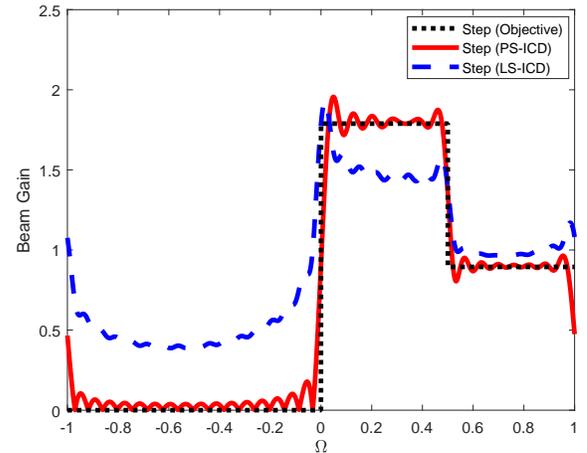}
\end{minipage}
}
 \caption{Comparison of triangular and step beam patterns for PS-ICD and LS-ICD.} \label{fig:VariousBeam}
\end{figure}

Fig.~\ref{fig:VariousBeam} extends our work to design more general beamforming vectors such as triangular beam and step beam with $N_{\rm t}=32$. Since  BMW-SS can not be used to design general beamforming vectors, we compare the beam patterns generated by PS-ICD and LS-ICD. Given the objective beam pattern, e.g., triangular beam or step beam, PS-ICD can better approach the objective beam than LS-ICD. Note that the triangular beam or step beam can be used for beam training by exploiting the overlapped beam patterns of neighbouring beams~\cite{IrregularWaveform}.

\begin{table}[!t]
\centering
\caption{Comparison of main lobe variation for different ideal codewords design methods}
\label{Tab.max deviation}
\begin{tabular}{ p{2.5cm}p{1.7cm}p{1.4cm}p{1.4cm}}
\toprule
Number of antennnas  &   PS-ICD       &   LS-ICD & BMW-SS  \\
\midrule
$N_{\rm t}=16$  &   $0.0015$  &      $0.0245$  &   $0.0425$    \\
$N_{\rm t}=32$  &   $7.834 \times 10^{-4}$  &    $0.0246$  &   $0.0160$   \\
$N_{\rm t}=64$  &   $4.939 \times 10^{-4}$  &    $0.0251$  &   $0.0160$   \\
$N_{\rm t}=128$ &  $3.435 \times 10^{-4}$  &     $0.0259$  &   $0.0140$    \\
\bottomrule
\end{tabular}
\end{table}

\begin{figure}[!t]
\centering
\includegraphics[width=88mm]{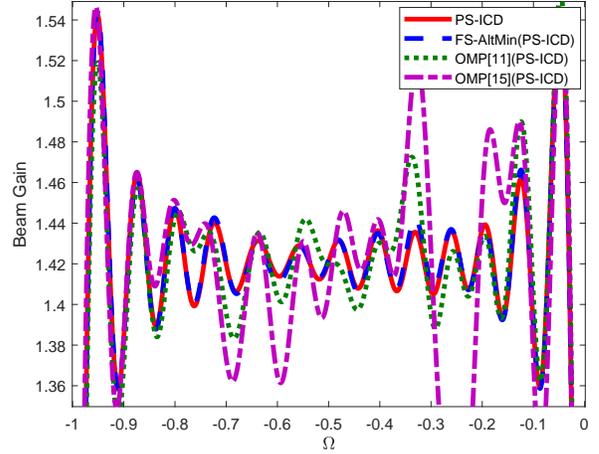}
\caption{Comparison of the beam patterns for different practical codeword design methods.} \label{fig:PDC}
\end{figure}

Fig.~\ref{fig:PDC} compares the beam patterns of practical codewords designed by FS-AltMin, OMP~\cite{CommonCodebookDesign} and OMP~\cite{Sparse2014}. To highlight the difference, we only illustrate the main lobe of codewords in Fig.~\ref{fig:PDC}. Since PS-ICD performs best among the ideal codeword design methods, we first generate ideal codewords using PS-ICD, where we set the parameters the same as Fig.~\ref{fig:beam pattern}, i.e., $N_{\rm t}=32$ and $\mathcal{I}_{v}=[-1,0]$. Then we generate practical codewords using FS-AltMin, OMP~\cite{CommonCodebookDesign} and OMP~\cite{Sparse2014} to approach the ideal codeword designed by PS-ICD. We set $b=6$ the same for FS-AltMin, OMP~\cite{CommonCodebookDesign} and OMP~\cite{Sparse2014}. The numbers of RF chains for FS-AltMin, OMP~\cite{CommonCodebookDesign} and OMP~\cite{Sparse2014} are $N_{\rm RF}=4$, $N_{\rm RF}=6$ and $N_{\rm RF}=15$, respectively. In addition, we set $T_{\rm max}=50$ for FS-AltMin. From Fig.~\ref{fig:PDC}, it is seen that FS-AltMin performs much better than OMP~\cite{CommonCodebookDesign} and OMP~\cite{Sparse2014}, which means FS-AltMin can better approach PS-ICD than OMP~\cite{CommonCodebookDesign} and OMP~\cite{Sparse2014}. Note that compared to FS-AltMin, OMP~\cite{Sparse2014} employs almost four times of the number of RF chains, which implies that FS-AltMin can save large hardware resource aside of generating better beam pattern.

As shown in Fig.~\ref{fig:Conver}, we compare the deviation between the designed practical codewords and the ideal codeword with different numbers of RF chains. Since PS-ICD has the best performance for {ideal codeword design}, an ideal codeword $\boldsymbol{v}$ is generated using PS-ICD, where we set the parameters the same as those of Fig.~\ref{fig:beam pattern}, i.e., $N_{\rm t}=32$ and $\mathcal{I}_{v}=[-1,0]$. Then we design a practical codeword $\boldsymbol{v}_{\rm p}$ with FS-AltMin, OMP~\cite{CommonCodebookDesign} and OMP~\cite{Sparse2014}. We define
\begin{equation}\label{Errorvandvp}
  E\triangleq\|\boldsymbol{v}-\boldsymbol{v}_{\rm p}\|_2
\end{equation}
to indicate the deviation between the ideal codeword $\boldsymbol{v}$ and the practical codeword $\boldsymbol{v}_{\rm p}$. We set $b=6$ the same for FS-AltMin, OMP~\cite{CommonCodebookDesign} and OMP~\cite{Sparse2014}. In addition, we set $T_{\rm max}=50$ for FS-AltMin. From Fig.~\ref{fig:Conver}, it is seen that FS-AltMin performs much better than OMP~\cite{CommonCodebookDesign} and OMP~\cite{Sparse2014}, which means with the same number of RF chains, the practical codeword designed by FS-AltMin has much smaller deviation than that designed by OMP~\cite{CommonCodebookDesign} or OMP~\cite{Sparse2014}.

{\color{red}
\begin{figure}[!t]
\centering
\includegraphics[width=88mm]{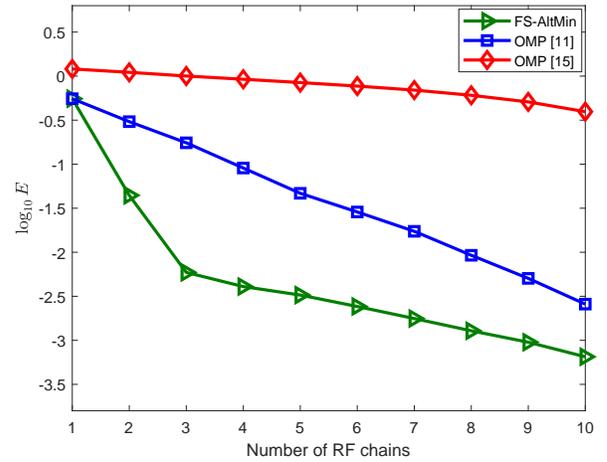}
\caption{Comparison of the deviation between the designed practical codeword and the ideal codeword for different numbers of RF chains.}
\label{fig:Conver}
\end{figure}
\begin{figure}[!t]
\centering
\includegraphics[width=88mm]{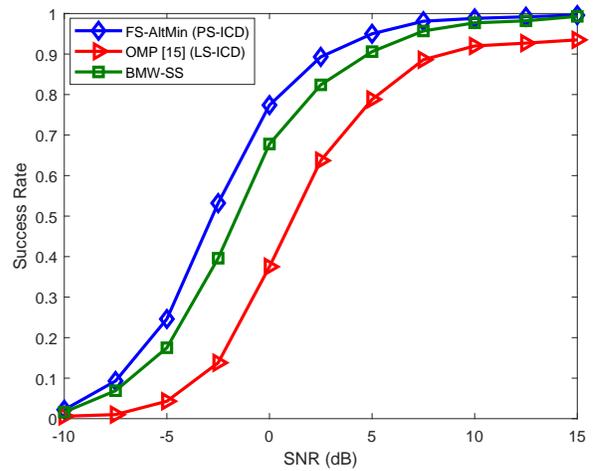}
\caption{Comparison of success rates of beam training using hierarchical codebooks designed by different methods.}
\label{fig:Codeword Compare}
\end{figure}
}

Fig.~\ref{fig:Codeword Compare} compares the success rates of beam training using the hierarchical codebooks, where the codewords in these codebooks are designed by three methods. The first two methods are all based on two-step codewords design, where PS-ICD and LS-ICD marked in the parenthesis in Fig.~\ref{fig:Codeword Compare} are used to design the ideal codewords. PS-ICD uses FS-AltMin to design the practical codewords with $N_{\rm RF}=4$ while LS-ICD uses OMP ~\cite{Sparse2014} to design the practical codewords with $N_{\rm RF}=12$. The last method is BMW-SS, which is designed using the overlapped adding of beams formed by sub-arrays and only considers the design of codewords. We set $N_{\rm r}=32$, $M=2$ and $L=1$ and perform the beam training algorithm based on the hierarchical codebooks proposed in~\cite{Sparse2014}. The definition of success rate is given in \eqref{BeamTrainingOptimizationProblem} and \eqref{BeamTrainingOptimizationProblem2}. In fact, the success rate in Fig.~\ref{fig:Codeword Compare} is determined by the ideal codewords as well as the practical codeword design methods, while the beam pattern in Fig.~\ref{fig:PDC} is only determined by the practical codeword design methods. From Fig.~\ref{fig:Codeword Compare}, it is seen that the proposed two-step codeword design method outperforms the existing methods. At signal-to-nose ratio (SNR) of 0dB, the proposed method has nearly 15\% improvement over BMW-SS. Note that the proposed method can be used to design codewords with arbitrary width of main lobe, which is different from BMW-SS.

\section{Conclusion}\label{Sec.Conclusion}
In this paper, we have investigated the codeword design for mmWave massive MIMO. A two-step codeword design approach has been developed. In the first step, additional phase is introduced to the beam gain to provide additional degree of freedom and the PS-ICD method has been proposed. To determine the additional phase, an alternative minimization method has been used, where each iteration of the method is based on a closed-form solution. The proposed PS-ICD can also be extended to design more general beamforming vectors with different beam patterns. Based on the ideal codewords designed in the first step, in the second step we have proposed a FS-AltMin algorithm that alternatively designs the analog precoder and digital precoder. In our future work, we will try incorporating some good merits of the FIR filter design, e.g. ~\cite{FIRDesign}, into the beam design in mmWave massive MIMO.

\bibliographystyle{IEEEtran}
\bibliography{IEEEabrv,IEEEexample}

 % that's all folks
\end{document}